# STUDY OF GAMMA INDUCED DEFECTS IN Nd DOPED PHOSPHATE GLASS USING UV-VIS SPECTROPHOTOMETER AND PHOTO-PHYSICS BEAMLINE ON INDUS-1

V. N. Rai[1], B. N. Raja Sekhar[2], B. N. Jagtap[2]

[1]Indus Synchrotron Utilization Division, Raja Ramanna Centre for Advanced Technology
Indore-452013, India

[2] Atomic and Molecular Physics Division, Bhabha Atomic Research Centre
Mumbai- 400085, India

Address for Correspondence
E-mail: vnrai@rrcat.gov.in
Phone: +91-731-2488142

# ABSTRACT

Nd doped phosphate glasses have been studied before and after gamma irradiation in order to understand the effect of glass composition and radiation induced defects on the optical properties of glasses. UV, Vis absorption and photoluminescence spectra of these glasses are found strongly dependent on the composition of glass matrix, particularly on the ratio of oxygen (O) and neodymium (Nd) concentration obtained from energy dispersive X-ray spectroscopic (EDX) measurement. Gamma irradiation of glass modifies the transmission below 700 nm due to generation of some new absorption bands corresponding to different types of defects. Observations indicate toward possibility of change in the valence state of $Nd^{3+}$ to $Nd^{2+}$ and generation of oxygen vacancies in glass matrix. EDX and X-ray photoelectron spectroscopic (XPS) measurements indicate change in the composition of glasses particularly decrease in the relative concentration of oxygen in glass samples after γ-irradiation.

# 1. INTRODUCTION

Optical properties of trivalent lanthanides have been extensively investigated in various host matrices particularly in different types of glasses. Study of glasses doped with lanthanide as well as transition metal ions have been found interesting due to its important optical properties, higher thermal expansion coefficients and lower transition temperature [1-8]. Glasses with different composition particularly phosphate glasses doped with lanthanide ions or transition metal ions have considerable potential in different field of applications such as solid state lasers, optical data transmission, solid state batteries, sensing laser technology, phosphors as well as for encapsulating the radioactive waste materials [9-10]. Phosphate glasses have received much attention because it can store much more optical density than other hosts and this energy can be efficiently extracted during the laser action. These glasses provide low refractive index and low dispersion. Preparation of phosphate glasses are relatively easy, where any change in its composition decides variation in its physical, chemical and optical properties, which may be of particular interest for specific technological applications. $Nd^{3+}$ doped phosphate glasses have been extensively studied for solid state lasers due to its laser emission at 1064 nm, in addition to the possibility of lasing at other wavelengths particularly at 1500, 1350 and 880 nm at room temperature [11]. The concentration of $Nd^{3+}$ and their structural groups present in the phosphate glass matrix depend on the percentage of modifiers, types and percentage of glass formers in the glass matrix, which decides the absorption and photoluminescence of the glass samples [12-13]. Even a systematic change in the composition of host material leads to substitutional changes in local structure, covalency and hence the related optical properties of doped rare earth ions in glasses. Studies related with glass structure and correlation of optical properties of Nd doped phosphate glasses with its elemental compositions are rather limited due to its hygroscopic nature and the volatility of $P_2O_5$ at elevated temperature [14].

It has been established that most of the glasses produce various interesting changes in their properties, when irradiated by high energy particles or radiation such as UV, γ rays and neutrons [15-24]. The irradiation effect produces changes in the optical properties of glasses in the form of generation of some new absorption bands. Recently a series of studies have been reported on γ-irradiation effect on different types of glasses such as soda lime phosphate, Cabal, bioglass, borosilicate, lithium borate and phosphate glasses [25-28], where intrinsic defects are found due to presence of dopants or impurity in the glass. Some of the dopants such as transition metal ions or lanthanide ions dopants capture negatively charged electrons or positively charged holes creating a change in their valence state through photochemical reaction during the exposure to successive gamma irradiation. However not much information is available on the effect of γ irradiation on Nd doped phosphate glasses. Effect of gamma irradiation on Nd doped gadolinium gallium garnet $Nd^{3+}:Gd_3Ga_5O_{12}$ (Nd:GGG) and $Nd^{3+}:Gd_3Sc_2Ga_3O_{12}$ (Nd:GSGG) laser crystals has been reported by Sun et al [20], which shows that GSGG crystal has stronger ability to resist the formation of color center than GGG. Zang et al [29] reported that $Yb^{3+}$ changes to $Yb^{2+}$ after γ irradiation in YAG crystal, which recovers after annealing. Similar changes in valency of $Nd^{3+}$ to $Nd^{2+}$ have been reported by Rai et al in neutron irradiated $Nd_2O_3$ powder and gamma irradiated Nd doped phosphate glasses.

This paper presents study of UV-Visible absorption, photoluminescence, EDX and XPS spectra of Nd doped phosphate glasses before and after γ- irradiation that provides a better understanding of the effect of glass composition and radiation induced defects on its optical properties. Results of measurements using conventional spectrophotometer and photo-physics beam line of Indian Synchrotron radiation source (INDUS-1) have been compared in order to evaluate the performance of this beamline for this type of study.

## 2. Experimental
### 2.1 Preparation of the glasses

The Nd doped phosphate glass samples (Obtained from CGCRI, Kolkata) were prepared using different composition of $P_2O_5$, $K_2O$, BaO, $Al_2O_3$, $AlF_3$, SrO and $Nd_2O_3$ as base materials. Different combination of base materials were used for getting four types of glass samples such as, sample #1 ($P_2O_5$, $K_2O$, BaO, $Al_2O_3$, $Nd_2O_3$), sample #2 ($P_2O_5$, $K_2O$, BaO, $Al_2O_3$, $Nd_2O_3$), sample #3 ($P_2O_5$, $K_2O$, BaO, $Al_2O_3$, $Nd_2O_3$, $AlF_3$) and sample #4 ($P_2O_5$, $K_2O$, SrO, $Al_2O_3$, $Nd_2O_3$). Their elemental compositions were measured after glass formation. Melt quenching technique was used to make these glasses, where reagents were thoroughly mixed in an agate mortar and placed in a platinum crucible for melting it in an electric furnace at $1095^0$ C for 1h 40 m. The melt was then poured onto a preheated brass plate and annealed at $365^0$ C for 18 h. Finally, the samples were polished to obtain smooth, transparent and uniform surface slab of 5 mm thickness for optical measurement. Energy dispersive X-ray spectra (EDX) of each samples were recorded before and after γ-irradiation using Bruker X-Flash SDD EDS detector, 129 eV in order to obtain the relative concentration of different elements (atomic %) present in the glass samples. It is necessary to find a correlation between the measured elemental concentration in the sample and the observed optical properties as well as to explain the experimental observations. Due to variation in concentration of elements from one place to other in the same sample, final data was obtained after averaging the data obtained at three random locations. Tables 1 to 3 show the average atomic % of the important elements present in the glass samples before and after gamma irradiation.

## 2.2 *γ- Irradiation of samples*

Few small pieces of glass samples were irradiated at room temperature using $^{60}$Co (2490 Ci, Gamma chamber 900) source of Gamma radiation having dose rate of 2 kGy/h. Samples were irradiated for radiation doses varying from 10 to 500 kGy.

## 2.3 *UV-Vis and photoluminescence spectroscopy*

Optical absorption spectra of the fresh and irradiated samples were recorded with a step size of 0.5 nm at room temperature in the range from 300 to 1100 nm using a spectrophotometer (Lambda 20 Perkin Elmer). Optical transmission spectra of these samples were also recorded using focused white light source from Indus-1 synchrotron radiation source. This experiment was performed using photo-physics beamline [30]. White light was obtained by setting the monochromator in zeroth order. This light from synchrotron was focused on the sample kept on a rotatable stage in the air and the light transmitted from the sample was collected using a lens coupled to an optical fiber which is connected to a spectrometer (OCEAN OPTICS - 2000). This spectrophotometer provides spectrum instantaneously and can provide average of many spectrum. The benefit of using this system for recording absorption spectrum is that it does not require scanning by grating thus reduces the recording time and is able to provide an averaged spectrum in very short time. It provides spectrum between 200-900 nm with a spectral resolution of 0.3 nm.

Polished samples of equal thickness ~ 5 mm were used for this measurement. Photo luminescence spectra of the samples were recorded before and after irradiation by a spectrofluorometer (Fluoromax-3 Yobin Yuan) using an excitation wavelength of 585 and 800 nm. The induced absorbance is used to describe the presence or generation of defects on the basis of difference spectra of the samples obtained from the spectra taken before and after irradiation.

The values of differential absorption or additional absorption (AA) spectra due to the irradiation are calculated using the formula [31] as

$$\Delta k = \frac{1}{d} \ln \frac{T_1}{T_2} \quad \text{-------------------------------------- (1)}$$

Where d is the sample thickness and $T_1$ and $T_2$ are the values of transmission intensity at particular wavelength obtained from the spectra of Nd doped phosphate glass recorded before and after γ-irradiation respectively. In fact $\Delta k$ is just a difference between linear absorption coefficient after and before irradiation, where bulk contribution cancels out and resulting spectrum remains mainly due to induced defects.

X-ray photoelectron spectra (XPS) of glass samples were also recorded before and after γ-irradiation using Al $K_\alpha$ line emission.

## 3. RESULTS AND DISCUSSION
### 3.1 *Optical (UV – Vis) absorption spectra*

The optical absorption spectra of $Nd^{3+}$ doped phosphate glasses having different composition (Table-1) were recorded in the wavelength range from 300-1100 nm as shown in Fig.-1. The absorption spectra recorded from all the four samples were analysed. The absorption transitions of $Nd^{3+}$ pertaining to $^4F_{3/2}$, $^4F_{5/2}$ + $^2H_{9/2}$, $^4S_{3/2}$ + $^4F_{7/2}$, $^4F_{9/2}$, $^2H_{11/2}$, $^4G_{5/2}$ + $^2G_{7/2}$, $^4G_{7/2}$, $^4G_{9/2}$, $^4G_{11/2}$ + $^2K_{15/2}$ + $^2G_{9/2}$ + $^2D_{3/2}$, $^2P_{1/2}$ + $^2D_{5/2}$ ← $^4I_{9/2}$ were observed, which is in agreement with the earlier reported results [3-4, 8]. The energies, oscillator strength and many other important parameters related to these transitions have been calculated from the absorption spectra of different Nd doped glass samples by many authors and have been used to explain the optical properties of these glasses on the basis of these calculated parameters [3-4, 8]. Present experiment shows that sample # 1 and # 3 have similar absorption amplitude for all the $Nd^{3+}$ peaks with higher absorption intensity. Sample # 2 shows higher background in lower wavelength side below 550 nm, whereas sample # 4 has higher background absorption through out the wavelength range. Intensity of the absorption peaks for sample # 4 is smaller but remains in between the absorption peak intensity observed for sample # 1 and # 2. Table - 1 shows that average atomic % of Nd in sample # 1 and # 3 is very close (0.32 and 0.28) and higher than sample # 2 and # 4 (0.09 and 0.19). Less amount of Nd present in sample # 2 and # 4 may be the reason for decreased peak intensity. However increase in background emission may be due to the combined effect of whole matrix in general, and particularly the concentration of main element O and Nd in the glass sample. The ratio of O to Nd obtained from Table-1 is very high for sample # 2 (914) and # 4 (409) in comparison to sample # 1 (243) and # 3 (289). This shows a simple correlation that the less value of O to Nd ratio provides good absorption in the glass, where as increase in the ratio increases the back ground absorption [32].

It has been reported [8] that in the case of Nd doping the spectral profiles of the absorption peaks in Lithium, lithium sodium, lithium potassium and sodium potassium phosphate glass matrices are similar. The spectral profiles of sodium, potassium and calcium phosphate glass matrices are also similar. It seems that mainly O and Nd concentrations are playing important role in deciding the absorption in Nd doped glass samples as O is the main element forming the covalent bonding with dopant and other elements in glass matrix. This is in agreement with the results of Seshadri et al [8] that position and intensity of certain electric dipole transitions of rare earth ions are very sensitive to the environment of the rare earth ions. These hypersensitive transitions have been found proportional to the covalency of (Rare Earth) RE-O bonds [8].

## 3.2 *Photoluminescence (radiative properties) of glasses*

The photoluminescence spectra of Nd doped phosphate glasses were recorded after exciting it with wavelength 584 and 800 nm to pump the metastable level $^4F_{3/2}$. In the present case attention has been paid to the photoluminescence emission band corresponding to $^4F_{3/2} \rightarrow {}^4I_{9/2}$ transition from all the four samples mainly for the purpose of comparison (Fig. 2). It is noted that sample # 1 and # 3 provide highest but nearly similar photoluminescence intensity. Sample # 4 provides medium intensity, whereas lowest intensity is found for sample # 2. Higher photoluminescence intensity is observed for the sample having higher absorption. This indicates that samples having higher background absorption and less photoluminescence may provide increased nonradiative transition. Decrease in photoluminescence for samples # 2 and # 4 seems to be due to presence of less amount of Nd and/or higher value of O to Nd ratio in comparison to sample # 1 and # 3 (Table-1).

The results of spectral absorption are utilized to understand the radiative properties of $Nd^{3+}$ ions. Normally three emission lines corresponding to the radiative transition of $Nd^{3+}$ ions from $^4F_{3/2} \rightarrow {}^4I_{9/2}, {}^4I_{11/2},$ and $^4I_{13/2}$ have been studied and reported by many authors [3, 4, 8-11]. Particularly $^4F_{3/2} \rightarrow {}^4I_{11/2}$ transition has been studied extensively, because it provides strong stimulated emission at 1.06 μm. The broad band emission centered at 880 nm ($^4F_{3/2} \rightarrow {}^4I_{9/2}$) has also been found important for Nd doped waveguide amplifiers operating with tunable wavelengths from 850 – 930 nm [33]. Presence of less amount of Nd (# 2 & # 4) indicates decrease in the peak absorption and photoluminescence intensity, where as increase in the ratio of O to Nd may be responsible for increased non-radiative transition. Even non-radiative de-excitation can occur if one of the excited rare earth ions is coupled with an –OH impurity [34]. It has also been reported [4,11] that the presence of fluoride content in glass samples increases the resistance to water and hence decreases the non-radiative relaxation of emitting level. The observed higher photoluminescence intensity in sample # 3 may be due to the presence of fluoride in stead of comparatively less amount of Nd (0.28 at. %) and higher value of O to Nd ratio (289) present in it (Table -1). This indicates that detailed information about the local environment of RE ions is important for understanding the absorption and photoluminescence behavior of RE containing glasses, which needs further investigations.

### 3.3 Gamma induced defects in glasses

The transmission (absorption) spectra of Nd doped phosphate glasses (Sample # 1 and 3) were recorded after gamma irradiation of 5-500 kGy at room temperature. Fig.-3 shows the transmission spectra of sample #1 irradiated for 5, 10 and 500 kGy along with unirradiated samples [24]. The unirradiated sample shows various peaks between 300 – 900 nm due to optical absorption of $Nd^{3+}$ ions present in the phosphate glass. It is noted that gamma irradiation of sample # 1 and 3 induces a broad absorption from 300 – 650 nm, which is similar for both the samples. An increase in the dose of γ-irradiation shows more decrease in transmission in the same wavelength range as is shown in Fig.-3. The $Nd^{3+}$ peaks at 750, 800 and 875 nm remain unaffected by the irradiation. In order to see the difference between pure glass and irradiated glass spectra, an additional absorption spectra or difference spectra (Δk) was obtained using eq.-1. Fig.-4a and 4b shows the difference spectra of sample #1 and 3 obtained after different doses of irradiation respectively. Fig.-4a shows that sample # 1 irradiated for 5 and 10 kGy has two positive peaks at 330 and 530 nm (broad peak) along with some negative peaks super imposed on it at the wavelength locations, where $Nd^{3+}$ has absorption peaks. A higher dose of (100 & 500 kGy) irradiation in sample # 1 increases the amplitude and broadening of absorption peaks in difference spectra. Finally two peaks (330 and 530 nm) observed separate at 10 kGy irradiation (sample #1) got merged and provided a broad band. Both the spectra from sample # 3 irradiated at 10 and 100 kGy (Fig.-4b) shows a single broad positive peak below 700 nm. However the negative peaks were observed similarly as in the case of sample # 1. The intensities of these broad peaks increase and negative peaks decrease with an increase in the dose of irradiation from 10 to 100 kGy. This observation is similar as seen for sample # 1 for gamma irradiation ≥ 100 kGy. This indicates that sample # 3 is more sensitive for gamma irradiation than sample #1. The negative peaks at 585, 750, 800 and 875 nm became more pronounced with an increase in dose of irradiation in both the samples (Fig.-4a and 4b). D. Sun et al [3] have also reported an increase in absorption in UV-Vis wavelength range in the case of Nd: GGG and Nd: GSGG crystal after gamma irradiation but the increase in absorption was less pronounced in comparison to Nd doped phosphate glasses. This indicates that glass samples are more sensitive under the effect of gamma radiation. Observation of new broad absorption bands below 700 nm in difference spectra of sample #1 and 3 indicates the creation of various kinds of defects in the glass sample after gamma irradiation. However, observation of negative peaks at the wavelength, where un-irradiated glass samples have peaks, indicates toward decrease in the number density of the $Nd^{3+}$ ions after irradiation. This decrease in number density of $Nd^{3+}$ ions and increase in the amplitude of new absorption bands in difference spectra are directly proportional to an increase in the dose of gamma irradiation.

Transmission spectrum of Nd doped phosphate glass before and after gamma irradiation was also recorded using white light from photo-physics beamline of Indus -1 synchrotron radiation source. Additional absorption spectrum was obtained for all the three irradiated (5 – 500 kGy) samples (Fig. - 5). These spectra also show some broad band absorption peaks between 300 – 585 nm. Three prominent negative peaks at 585, 750 and 800 nm are also observed similar as shown in Fig.– 4a and 4b obtained using spectrophotometer. However other absorption peaks of $Nd^{3+}$ observed before irradiation between 300 to 585 nm are also observed in the form of negative peaks superimposed on the new broad additional absorption bands obtained from Indus-1 light source (Fig.-5). This indicates that better resolution of OCEAN OPTICS spectrometer and intense synchrotron radiation make it possible to record very week absorption bands, which were not possible with spectrophotometer. Another advantage of this technique is that it records whole spectrum instantaneously and can provide average of a number (128) of spectra, which is useful in reducing the random noise in the spectrum.

For making the confirmation of changes in valency of $Nd^{3+}$ to $Nd^{2+}$ as a result of gamma irradiation, photoluminescence spectra of Nd doped phosphate glasses were recorded and compared before and after gamma irradiation using excitation wavelength of 585 nm, which is a highly absorbing line of $Nd^{3+}$. Fig. - 6 shows that unirradiated phosphate glass has very intense photoluminescence peak centered at 805 and 870 nm and a small peak centered at 745 nm. The irradiated sample shows a low intensity peak shifted to higher wavelength side near 825 and 875. Similar decrease in luminescence of $Nd^{3+}$ in GGG crystal has been reported by Sun et al [3]. This decrease in luminescence intensity also indicates towards possible decrease in number density of $Nd^{3+}$ ions. This observation confirms that gamma irradiation changes the valence state of Nd ions from $Nd^{3+}$ to $Nd^{2+}$ along with an increase in other types of defects.

Radiation induced damage in different phosphate glasses have been reviewed by Lell et al [35], Bishay [36] as well as Friebele and Griscom [37]. Bishay [36] has reported that radiation induced optical absorption has been observed from 200 – 540 nm, where 540 and 420 nm bands may be due to positive hole trap center in the glass, where as $O^{2-}$ vacancies can be electron trapping site. Moencke and Ehrt [38] have identified many irradiation induced defects in a series of phosphate glasses. They assume that radiation induced defects in glasses are formed in pairs of negative electrons centers and positive hole centers. F. H. ElBatal et al [21] have studied gamma irradiation effects on $V_2O_5$ doped sodium phosphate glass and reported various peaks in additional absorption spectrum at 230, 330, 375, 405 and a broad band at 485nm. The band at 485 nm showed prominent growth with gamma dose. It is accepted that when a glass/crystal containing transition or rare earth ions are irradiated, these ions act as a potential trap for the released electrons and holes during the irradiation process. Trapping of charges by these ions are favoured and the behavior of such material (glass/crystal) depends on the type and concentration of these dopants or impurities. Present experimental observations as decrease in number density of $Nd^{3+}$ ions and increase in amplitude of absorption band below 700 nm (in difference spectrum) indicates that $Nd^{3+}$ may be changing its valency after capturing electron or hole created under the effect of gamma irradiation. For identification of the possible changes, absorption spectra of $Nd^{2+}$ in different matrix were compared with the additional absorption spectra of Nd doped glass sample. It has already been reported that the energy levels of $Nd^{2+}$ ions in single crystal $CaF_2$ [39] show an absorption band centered at 527 nm due to 4f – 5d transition of $Nd^{2+}$ ions. Similar absorption band has also been observed by Xu and Peterson in Nd doped $SrB_4O_7$ matrix [40]. Rai et al. [16] have reported possible conversion of $Er^{3+}$ and $Nd^{3+}$ to $Er^{2+}$ and $Nd^{2+}$ in $Er_2O_3$ and $Nd_2O_3$ powder respectively under the effect of neutron irradiation. Similar change in valency from $Yb^{3+}$ to $Yb^{2+}$ has been reported by X. Zeng et al [29] after gamma irradiation of Yb: YAG crystal. The comparison suggests that broad peak observed in the difference spectra of irradiated Nd doped phosphate glass may be having some amount of $Nd^{2+}$, which indicates conversion of $Nd^{3+}$ to $Nd^{2+}$ under the effect of gamma irradiation.

Other types of defects are also generated in the glass material after the gamma irradiation [18-29] such as oxygen vacancies and color centers. EDX measurements on gamma irradiated sample #1 and 3 (Table-2 an3) support this idea and shows that composition of glasses change after gamma irradiation. Major change was noted as decrease in concentration of oxygen after gamma irradiation of 500 kGy. At lower doses of irradiation these changes are not resolvable. The generation of $O^-$ center after gamma irradiation in different doped crystal has been studied by many authors. Mares et al [41] suggested about an additional absorption in the range 360 – 480 nm (peaking at 380 – 425 nm), which was attributed to the absorption of $O^-$ centers in gamma irradiated Ce: LuYAP crystal. Kim et al [42] also reported the absorption peaks at 428 nm due to absorption of $O^-$ centers in $PbWO_4$ crystal. Zeng et al [43] found absorption at 390 nm due to $O^-$ centers in Ce: YAP crystal. Ebeling et al [44] reported the presence of phosphorus and oxygen related radiation induced defects in phosphate glasses such as phosphorus oxygen hole center (POHC) absorbing in visible spectral region at 540, 430 and 325 nm. In the light of above discussion one can say that contribution from all types of defects ($O_2^-$, change in valency of ions and other color centers) overlap and show increase in amplitude and bandwidth of the absorption band ranging from 300 – 600 nm depending on the dose of irradiation.

### 3.4 Process of defect generation

Generally high energy gamma irradiation produces charged defects in the material very easily. It seems that when photon passes through the phosphate glass, it interacts with the ions present in the material. Possibility of Compton Effect is more when the energy of photon is more that 0.1 MeV. During Compton scattering part of the energy of photon is transferred to Compton electron detached from atom, which again interact with other electrons in the material and produces number of secondary electrons. These secondary electrons can be captured by the oxygen vacancies as well as by the other ions particularly $Nd^{3+}$ ions in the present experiment. This way $Nd^{3+}$ ion will change to $Nd^{2+}$, which is observed in the experimental results in the form of decrease in number density of $Nd^{3+}$ ions and increase in the intensity due to increase in the number density of other defects (Oxygen vacancies and other color centers). These processes can be expressed as

$$O^{2-} \text{ (collided by Compton electron)} \rightarrow O^{-} + e^{-}$$
$$V_O + e^{-} \rightarrow F^{+} \text{ (one electron trapped at vacancies of oxygen ion)}$$
$$F^{+} + e^{-} \rightarrow F \text{ (Two electrons trapped at the vacancy of oxygen ion)}$$
$$Nd^{3+} + e^{-} \rightarrow Nd^{2+}$$

These irradiation induced changes seem to be either irreversible or very slowly changing at room temperature in the present experiment as it provided similar optical behavior even after 6 months. This may be due to non conducting nature of material, where mobility of defects is very small or negligible. In the case of neutron irradiation of $Nd_2O_3$ powder, irradiation effect was recovered within few days at room temperature [16]. Generally reversal in the process has been observed in some of irradiated crystal after annealing [29].

### 3.5 *Effect of γ-irradiation on XPS of phosphate glass*

X- Ray photoelectron spectra (XPS) of Nd doped phosphate glasses were recorded before and after γ-irradiation using monochromatic Al $K_\alpha$ line emission. X-ray was focused on the surface of the sample. Core level of Nd 3d spectrum was studied. Fig.-7 shows the XPS spectra of glass sample # 1 before and after 10 & 500 kGy γ-irradiation. The inbox shows two broad peaks at 982.5 and 1003.7 eV due to $3d_{5/2}$ and $3d_{3/2}$ (Nd 3d spectra). These peaks become sharper after gamma irradiation and shifts to higher kinetic energy. Observation of sharpening and shift in the main peak towards higher energy side seems to be due to decrease in oxygen content in the glass sample after γ-irradiation. It is in agreement with the EDX data, which also shows that after γ-irradiation (500 kGy) at. % of oxygen content in sample # 1 decreases (Table-2). This decrease in the concentration of oxygen in EDX data is not observable for low dose of irradiation as 10 kGy. Fig.-8 shows that sample #1 & # 3 also have sharp Nd 3d peak due to $3d_{5/2}$ after γ irradiation of 10 kGy. It is observed that main peak of sample # 3 shifts to lower energy side and is broad in comparison to sample # 1, which may be due to the presence of $AlF_3$ and higher concentration of oxygen in sample # 3 network. The chemical shift or shift in peak position due to change in oxidation state of Nd is not clear.

XPS and EDX results indicate that concentration of oxygen decreases after γ-irradiation of glass samples. Similar variations in XPS of Nd containing alloy glass have been reported by Tanaka et al [45]. They have shown that the presence of higher concentration of oxygen in glass induces broadening in the peaks of $3d_{5/2}$ and $3d_{3/2}$ due to Auger O KLL and 3d satellite peaks. An O KLL peak represents the energy of the electrons ejected from the atoms due to the filling of the O 1s state (K shell) by an electron from the L shell coupled with the ejection of an electron from an L shell. This broadening due to presence of extra peaks is not observed in the sample, where oxygen content is less. The decrease in oxygen content is accompanied by a small shift in the main peak (Nd 3d) towards higher energy. Similar decrease in oxygen concentration after electron beam irradiation in glass samples has been reported by Puglisi et al [46]. They have used XPS to study the compositional changes in glass sample after electron beam irradiation and reported an out gassing of oxygen from the sample. XPS is effective mainly on surfaces as X-rays can not penetrate the glass sample. Therefore, it can not provide information about the bulk sample. The charging of the sample surface during XPS measurement may also affect the information about the charge state of Nd.

## 4. CONCLUSION

UV, Visible and photoluminescence spectra of Nd doped phosphate glasses have been studied before and after γ-irradiation. It has been found that results obtained from photo-physics beamline are satisfactory. Higher absorption and low background were observed in UV- visible spectra of glass samples (due to $Nd^{3+}$ ions) mainly for lower value of O to Nd concentration ratio. Similarly samples having higher absorption (low O to Nd ratio) result in higher photoluminescence. Gamma irradiation on phosphate glass produces various types of defect. The possible defects responsible for broadening in difference spectra seems to be due to the change in valence state of $Nd^{3+}$ to $Nd^{2+}$ along with generation of some other defects as $O^-$ vacancies, POHC and F centers. The decrease in oxygen content of glass sample after γ-irradiation as indicated by XPS and EDX measurements also confirms the generation of defects due to oxygen vacancy. Results also show that sample # 3 is soft for gamma irradiation than sample # 1. These changes are found dependent on the composition of the glasses as well as on the doses of irradiation.

Such studies may be helpful in finding glasses with better radiative properties for high power lasers as well as radiation hard glass materials. The decrease in transmission to 0% in Nd doped phosphate glass between 300 – 585 nm after gamma irradiation with a dose of 500 kGy as well as its irreversible nature can find use in many applications. Such a material can work as an optical filter for wavelength range 300 – 585 nm particularly for cutting off the second and third harmonics of Nd: YAG and Glass laser (1 µm). This process may be useful in the space applications or in similar conditions, where radiation level is so high, because in that condition unirradiated Nd doped phosphate glass will change its optical properties as filter, which avoids pre-irradiation of Nd doped phosphate glass before using it as a filter. A further investigation is needed to better understand the local environment of rare earth ions in glass matrix using different experimental techniques such as EXAFS.

**Acknowledgment**

Authors are thankful to S. Kher, P. Tiwari and D. M. Phase for their help during this experiment. Support and encouragement provided by S. K. Deb is also gratefully acknowledged.

**REFERENCES**

[1]. K. Gatterer, G. Pucker, W. Jantscher, H. P. Fritzer, S. Arafa, J. Non-Cryst. Solids **281** (1998) 189.

[2] E. Metwalli, M. Karabulut, D. L. Sideborton, M. M. Morsi, R. K. Brow, J. Non-Cryst. Solids **344** (2004) 128.

[3] B. Karthikeyan, S. Mohan, Mat. Res. Bull. **39** (2004) 1507.

[4] C. K. Jayasankar, R. Balakrishnaiah, V. Venkatramu, A. S. Joshi, A. Speghini, M. Bettinetti, J. Alloy and Compounds **451** (2008) 697.

[5] T. Minami, K. I. Mazawa, M. Tanaka, J. Non-Cryst. Solids **42** (1980) 469.

[6] M. Jamnicky, P. Znasik, D. Tureger, M. D. Ingram, J. Non-Cryst. Solids **185** (1995) 151.

[7] N. Satyanarayana, G. Gorindoraj, A. Karthikeyan, J. Non-Cryst. Solids **136** (1999) 219.

[8] M. Seshadri, K. Venkata Rao, J. Lakshamana Rao, K. S. R. Koteswara Rao and Y. C. Ratnakaran, J. Lumin. **130** (2010) 536.

[9] D. E. Day, Z. Wu, C. S. Ray, P. Hrma, J. Non-Cryst. Solids **241** (1998) 1.

[10] M. Karabulut, G. K. Marasinghe, C. S. Ray, D. E. Day, G. D. Waill, C. H. Booth, P. G. Allen, J. J. Bucker, D. L. Caulder, D. K. Shuh, J. Non.-Cryst. Solids **306** (2002) 182.

[11] R. Balakrishnaiah, P. Babu, C. K. Jayasankar, A. S. Joshi, A. Speghini and M. Bettinelle, J. Phys.: Condens. Matter **18** (2006) 165.

[12] J. H. Campbell and T. I. Suratwala, J. Non-Cryst. Solids **263/264** (2000) 318.

[13] M. Ajroud, M. Haouari, H. Ben Ouada, H. Maaref, A. Brenier and C. Garapon 2000 J. Phys.:Condens. Matter **12** (2000) 3181.

[14] R. Hussin, D. Holland, R. Dupree, J. Non.-Cryst. Solids **298** (2002) 32.

[15] E. J. Friebele, Radiation effects on Optical Properties of Glass, Ed. D. R. Uhlmann and N. J. Kreidl, American Ceramic Society, Westerville OH PP 205-260 (1991).

[16] V. N. Rai, S. N. Thakur and D. K. Rai, Appl. Spectrosc. **40** (1986) 1211.

[17] G. Sharma, K. Singh, Manupriya, S. Mohan, H. Singh and S. Bindra, Rad. Phys. And Chem. **75** (2006) 959.

[18]   F. H. ElBatal, Nucl. Instrum. And Methods in Phys. Res. B **265** (2007) 521.

[19]   A. K. Sandhu, S. Singh and O. P. Pandey, J. Phys D, Appl. Phys. **41** (2008) 165402.

[20]   D. Sun, J. Luo, Q. Zhang, J. Xiao, J. Xu, H. Jiang and S. Yin, J. of Liminescence **128** (2008) 1886.

[21]   F. H. ElBatal, Y. M. Hamdy and S. Y. Marzouk, Matr. Chem. Phys. **112** (2008) 991.

[22]   S. M. Abo-Naf, M. S. El-Amiry and A. A. Abdul-Khalik, Opt. Mater. **30** (2008) 900.

[23]   F. H. ElBatal, M. A. Ouis, R. M. M. Morsi and S. Y. Marzouk, J. Non-Cryst. Solids **356** (2010) 46.

[24]   V. N. Rai, B. N. Raj Shekhar, S. Kher and S. K. Deb, J. Limin. **130** (2010) 582.

[25]   F. H. ElBatal, S. M. Abo-Naf, F. M. EzzElDin, Indian J. Pure Appl. Phys. **43** (2005) 579.

[26]   S. Y. Marzouk, F. H. ElBatal, H. M Salem, S. M. Abo-Naf, Opt. Mater. **29** (2007) 1456.

[27]   F. H. ElBatal, M. A. Azooz, S. Y. Marzouk, Phys. Chem. Glasses, Eur. J. Glass Sci. Technol. **47** (2006) 588.

[28]   F. H. ElBatal, A. A. Elkheshn, M. A. Azooz, S. M. Abo-Naf, Opt. Mater **30** (2008) 881.

[29]   X. Zeng, X. Xu, X. Wang, Z. Zhao, G. Zhao and J. Xu, Spectrochim Acta Part A **69** (2008) 860.

[30]   N. C. Das, B. N. Raja Sekhar, S. Padmanabhan, S. Aparna, S. N. Jha, S. S. Bhattacharya, J. Opt. **32** (2003) 189.

[31]   A. Matkovski, A. Durygin, A. Suchoki, D. Sugak, G. Neuroth, F. Wallrafen and V. Grabovski, Opt. Mater. **12** (1999) 75.

[32]   V. N. Rai, B. N. Raja Sekhar, P. Tiwari, R. J. Kshirsagar and S. K. Deb, J. Non Cryst. Solids **357** (2011) 3757.

[33]   K. Kuriki and Y. Koike, Chem. Rev. **102** (2002) 2347.

[34]   Y. Yan, A. J. Faber, H.de Waal, J. Non-Cryst. Solids **181** (1995) 283.

[35] E. Lell, N. J. Kreidl, J. R. Hensler and J. Burke (Ed.) Progress in Ceramic Science Vol. **4** Pergamon Press, New York, 1966.

[36] A. Bhishay, J. Non-Cryst. Solids **3** (1970) 54.

[37] E. J. Friebele, D. L. Griscom, in M. Tomozawa, R. H. Doremus (Ed) Treatise on Material Science and Technology, Vol. **17** Academic Press, NewYork (1979).

[38] D. Moncke and D. Ehrt Opt. Mater. **25** (2004) 425.

[39] D. S. Mc Clure and Z. Kiss, J. Chem. Phys. **39** (1963) 3251.

[40] W. Up and J. R. Peterson, J. Alloy and Compounds **249** (1997) 213.

[41] J. A. Mares, N. Chekhov, M. Nike, J. Vail, R. Katy, and J. Pepsis, J. Alloy Comp.. **200** (1998) 275.

[42] T. H. Kim, S. Cho, K. Lee, M. S. Jang and J. H. Ro, Appl. Phys. Lett. **81** (2002) 3756.

[43] X. Zeng, G. Zhao, J. Xu, X. He, J. Appl. Phys. **95** (2004) 749.

[44] P. Ebeling, D. Ehrt and M. Friedrich, Optical Materials **20** (2002) 101.

[45] K. Tanaka, M. Yamada, T. Okamoto, Y. Narita and H. Takaki, Mat. Sci. Eng. **A181/A182** (1994) 932.

[46] O. Puglisi, G. Marletta and A. Torrisi, J. Non Cryst. Solids **55** 433 (1983).

**FIGURE CAPTION**

1. Absorption spectra (UV-Vis) of Nd doped phosphate glasses, (1) Sample #1 ($P_2O_5$,$K_2O$,BaO,$Al_2O_3$,$Nd_2O_3$) O/Nd=243.25, (2) Sample #2 ($P_2O_5$,$K_2O$,BaO,$Al_2O_3$,$Nd_2O_3$) O/Nd = 914.66, (3) Sample #3 ($P_2O_5$,$K_2O$,BaO,$Al_2O_3$,$Nd_2O_3$,$AlF_3$) O/Nd = 289.5, (4) Sample # 4 ($P_2O_5$, $K_2O$,SrO,$Al_2O_3$,$Nd_2O_3$) O/Nd = 409.84,

2. Photoluminescence spectra of Nd doped phosphate glasses recorded at excitation wavelength of 800 nm, (1) Sample #1, (2) Sample #2, (3) Sample #3, (4) Sample # 4.

3. Transmission spectra of Nd doped phosphate glass (1) Normal (unirradiated), (2) Gamma irradiated 5 kGy, (3) Gamma irradiated 10 k Gy, (4) gamma irradiated 500 kGy

4a. Differential or additional absorption spectra of gamma irradiated Nd doped phosphate glass (A) 500 kGy, (B) 10 kGy, (C) 5 kGy

4b. Differential absorption spectra (UV-Vis) of Nd phosphate glass Sample #3 obtained using eq.-1, (1) After γ-irradiation of 10 kGy, (2) After γ-irradiation of 100 kGy.

5. Photo luminescence of Nd doped phosphate glass (1) Un irradiated, (2) gamma irradiated 10kGy.

6. Differential absorption spectra of gamma irradiated Nd doped phosphate glass recorded using white light source from Indus – 1 synchrotron radiation source (A) 500 kGy, (B) 10 kGy, (C) 5 kGy.

7. XPS spectra of sample # 1 before and after γ-irradiation, (1) Pure sample #1 (2) Sample #1 irradiated at 500 kGy, (3) Sample #1 irradiated for 10 kGy

8. XPS spectra of glass samples after γ-irradiation of 10 kGy, (1) Sample #1, (2)Sample #3

Table – 1 Atomic % of elements present in the glass (EDX data)

| S. No. | Atomic Component | Sam. #1 (At. %) | Sam. #2 (At. %) | Sam. #3 (At. %) | Sam.#4 (At. %) |
|---|---|---|---|---|---|
| 1. | O | 77.84±6.35 | 82.32±5.50 | 81.06±5.60 | 77.87±7.47 |
| 2. | P | 13.20±0.60 | 10.27±0.43 | 9.99±0.43 | 13.64±0.73 |
| 3. | Ba | 1.94±0.45 | 1.32±0.26 | 1.77±0.37 | ———— |
| 4. | K | 3.96±0.25 | 3.58±0.16 | 4.73±0.20 | 3.19±0.20 |
| 5. | Al | 2.27±0.10 | 2.00±0.13 | 1.97±0.10 | 3.99±0.30 |
| 6. | Nd | 0.32±0.13 | 0.09±0.10 | 0.28±0.10 | 0.19±0.10 |

Table – 2 Atomic % of elements present in Sample #1 before and after γ irradiation (EDX data)

| At. Comp. | O | P | Ba | K | Al | Nd |
|---|---|---|---|---|---|---|
| Av. At. % | 77.84 ±6.35 | 13.20 ±0.60 | 1.94 ±0.45 | 3.96 ±0.25 | 2.27 ±0.10 | 0.32 ±0.13 |
| Av. At. % (γ irradiated 10 kGy) | 75.45 ±4.40 | 12.86 ±0.50 | 2.38 ±0.30 | 5.27 ±0.2 | 3.08 ±0.10 | 0.30 ±0.10 |
| Av. At. Wt. % (γ irradiated 500 kGy) | 66.19 ±4.3 | 17.00 ±0.60 | 3.12 ±0.45 | 5.90 ±0.25 | 3.99 ±0.15 | 0.37 ±0.10 |

Table-3 Atomic % of elements present in the sample #3 before and after γ irradiation (EDX data)

| At. Component | O | P | Ba | K | Al | Nd |
|---|---|---|---|---|---|---|
| Av.At.% | 81.06 ±5.60 | 9.99 ±0.43 | 1.77 ±0.37 | 4.73 ±0.20 | 1.97 ±0.10 | 0.28 ±0.10 |
| Av. At. % (γ irradiated 10 kGy) | 82.83 ±8.40 | 7.21 ±0.40 | 0.49 ±0.20 | 0.99 ±0.10 | 3.85 ±0.25 | 0.34 ±0.15 |

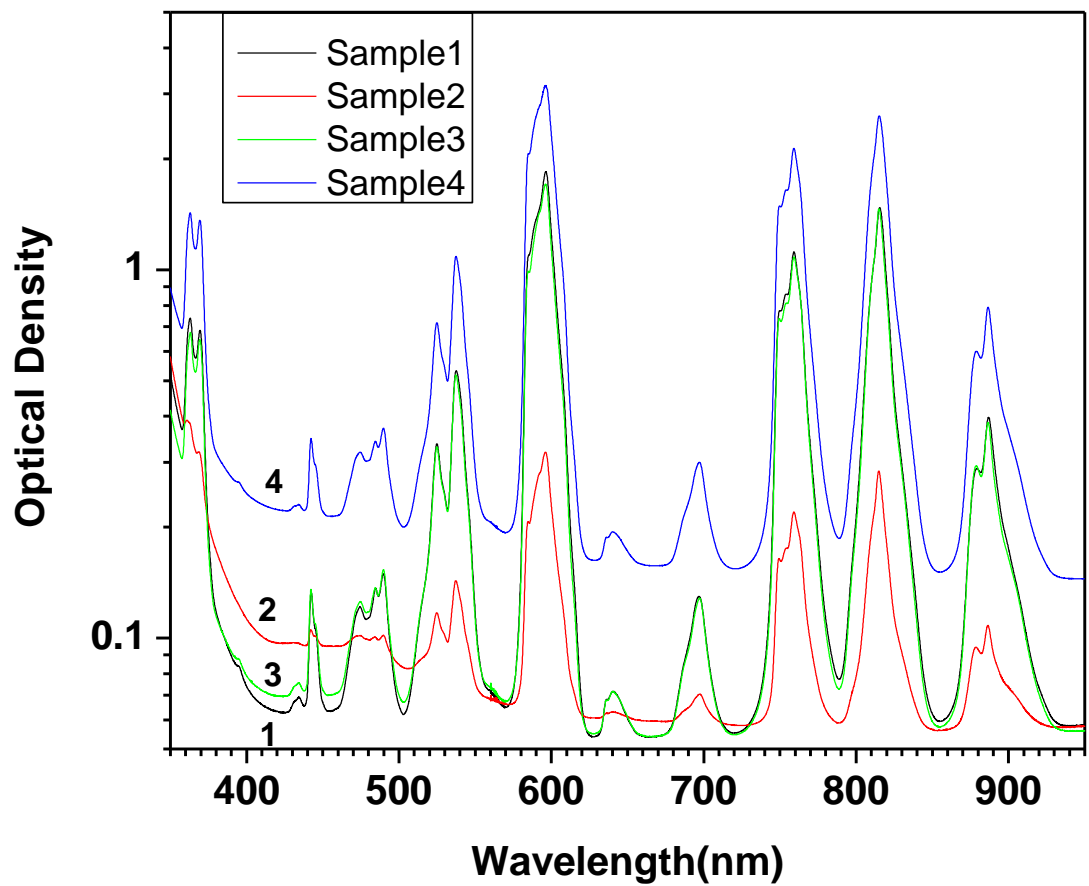

**Fig.-1**

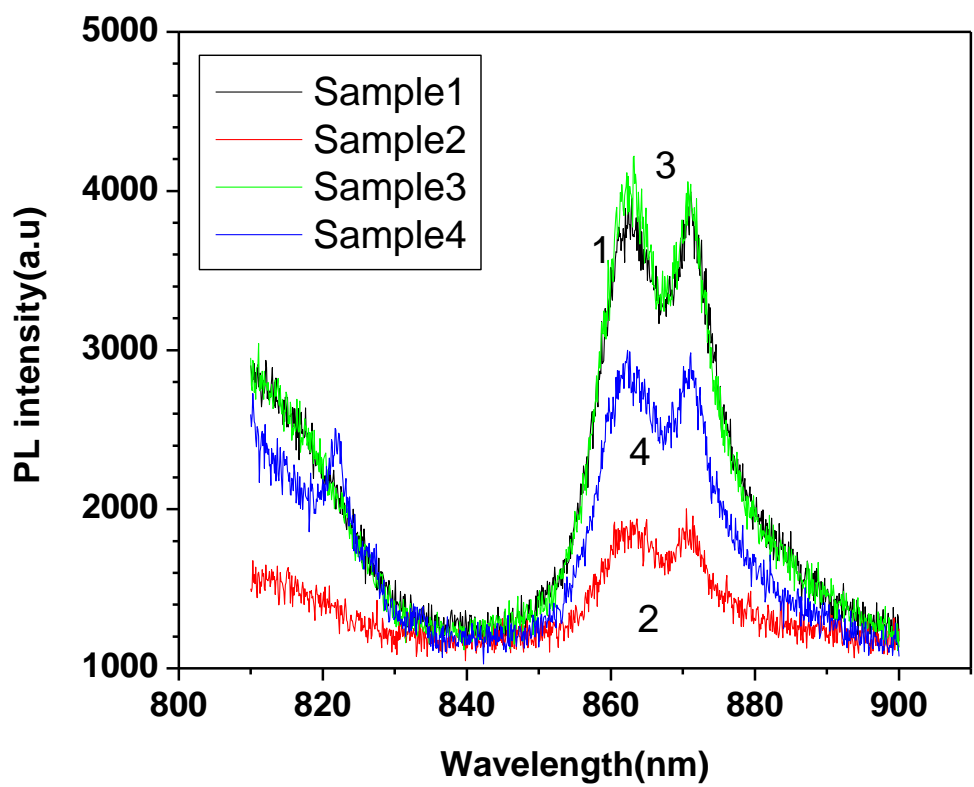

**Fig.-2**

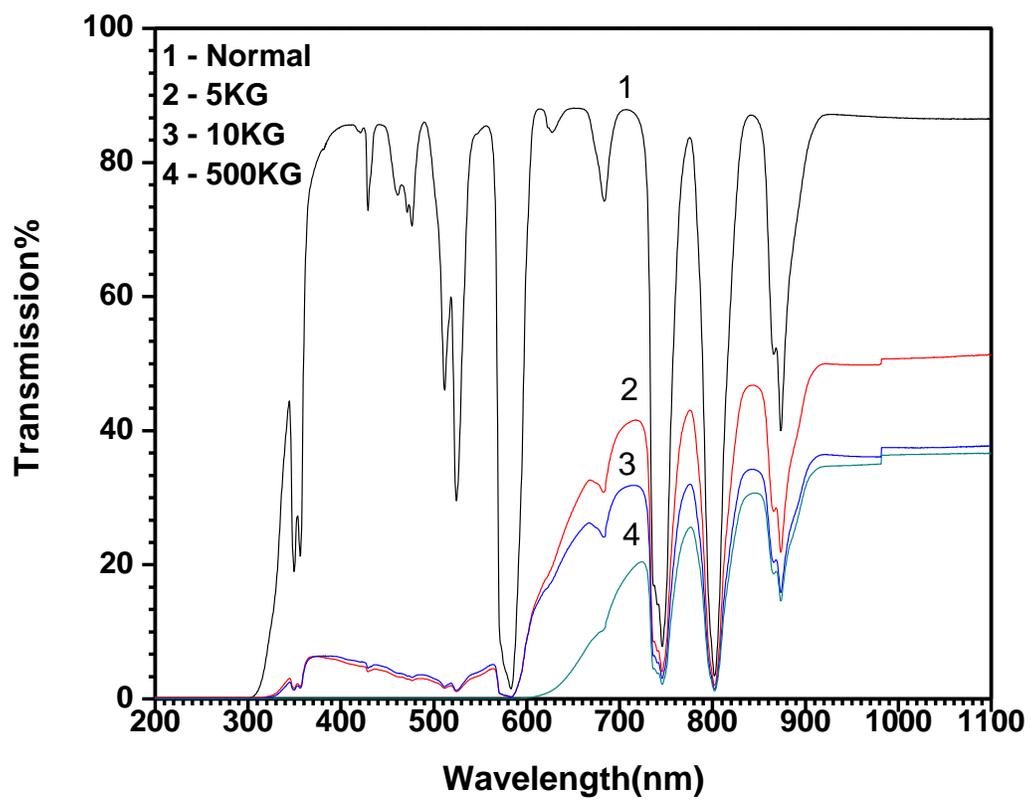

**Fig.-3**

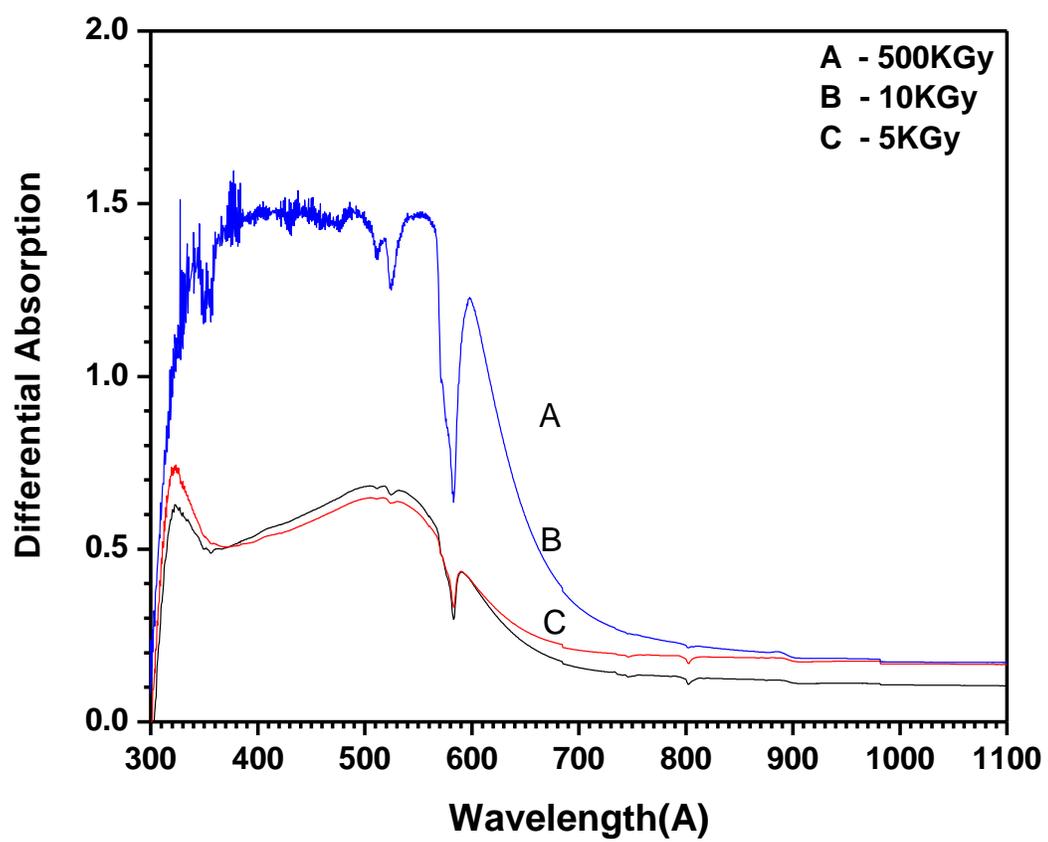

**Fig.-4a**

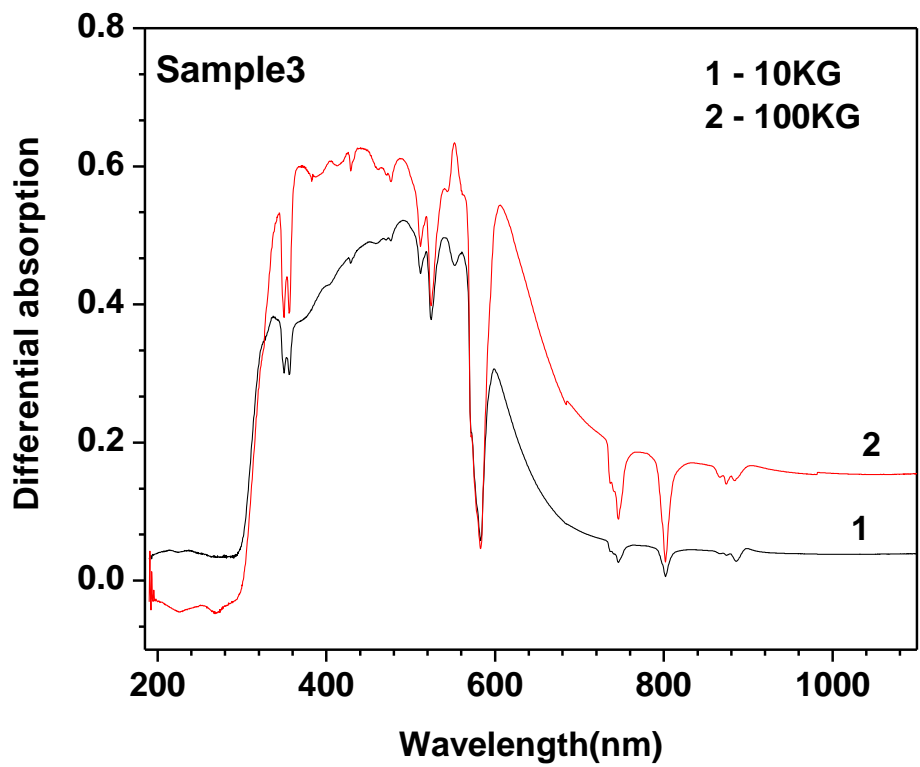

**Fig.-4b**

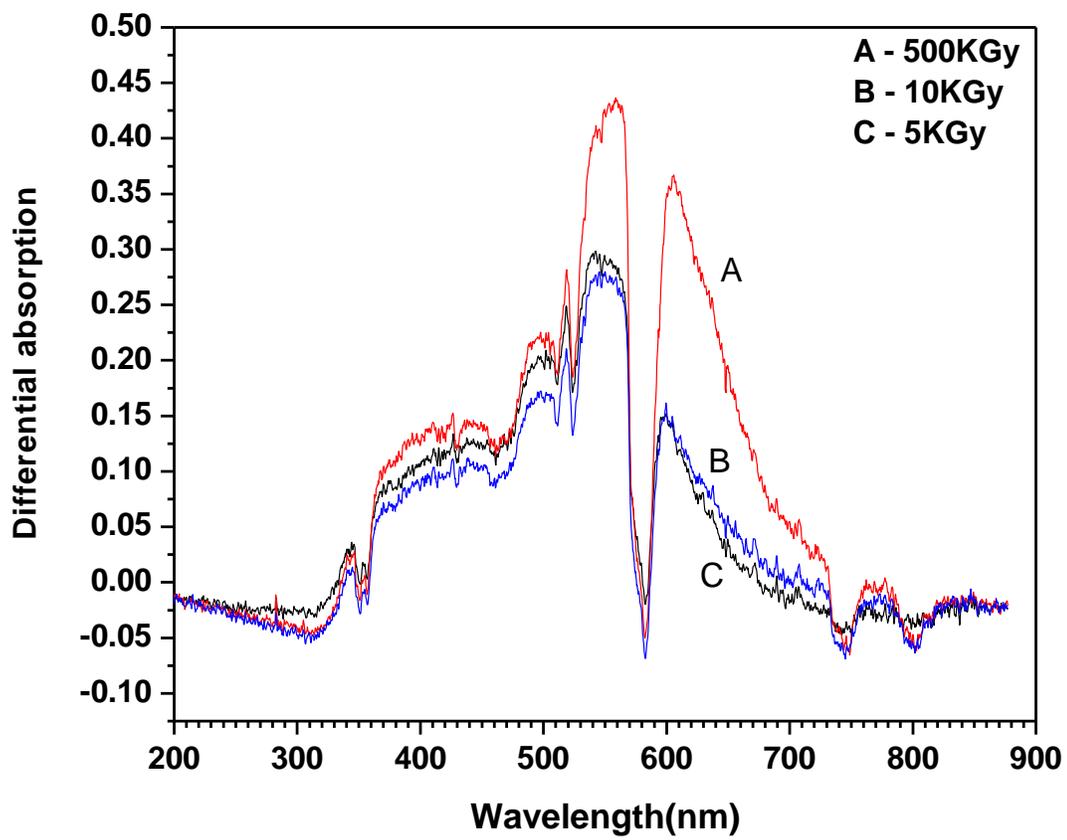

**Fig.-5**

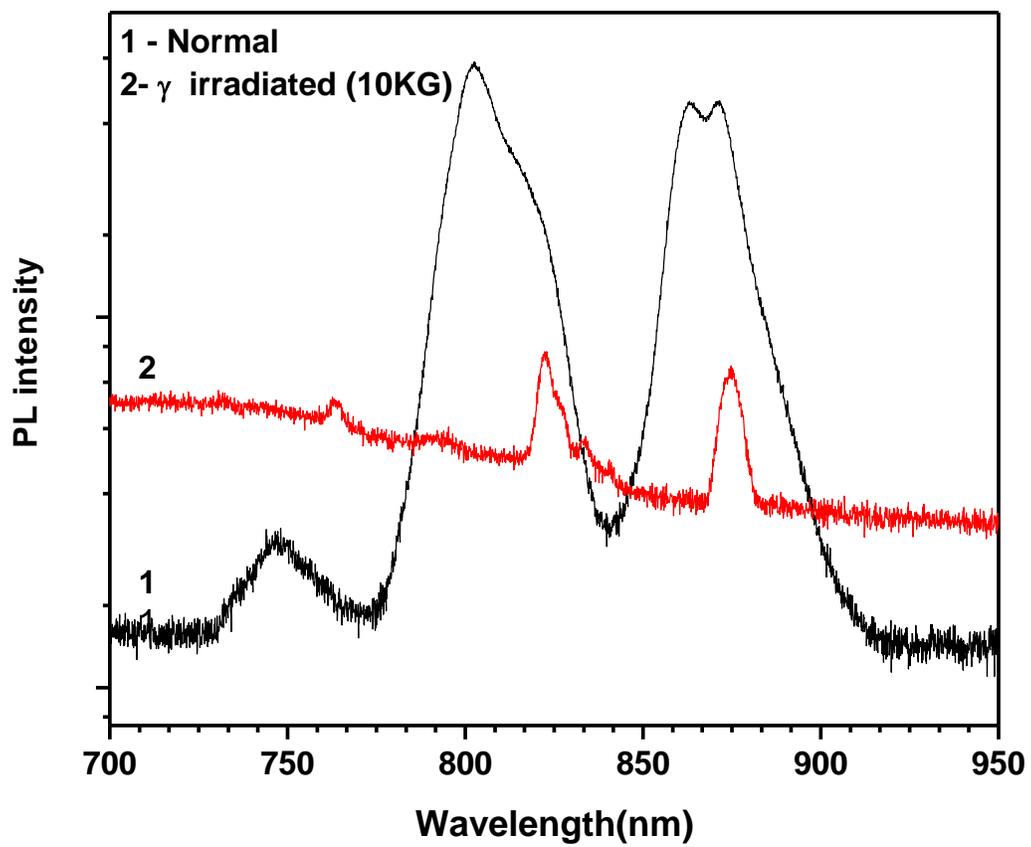

**Fig.-6**

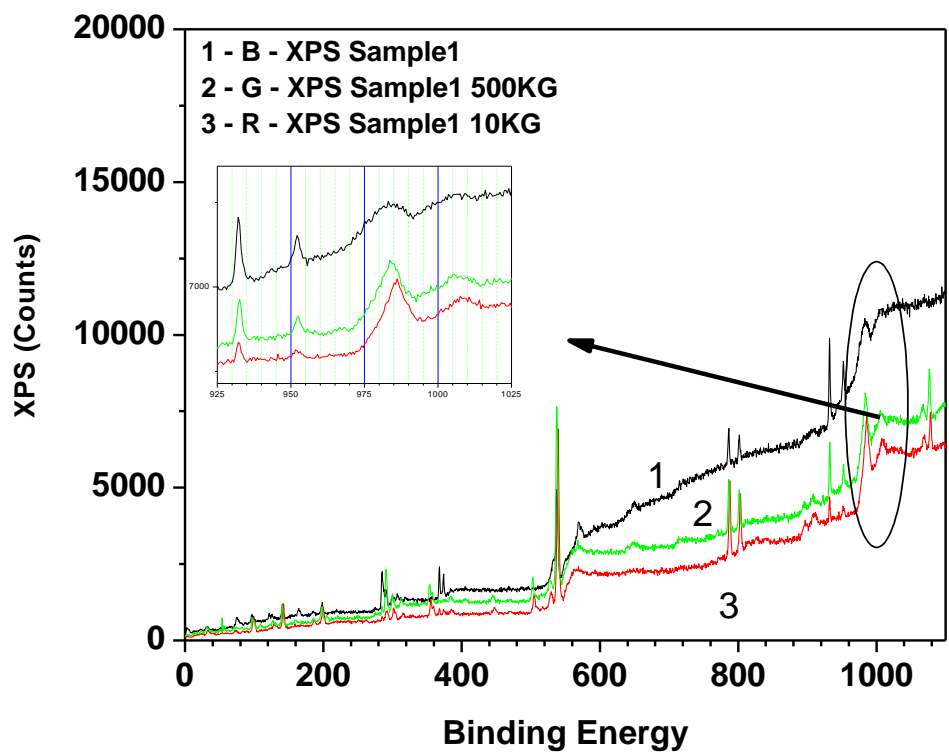

**Fig.-7**

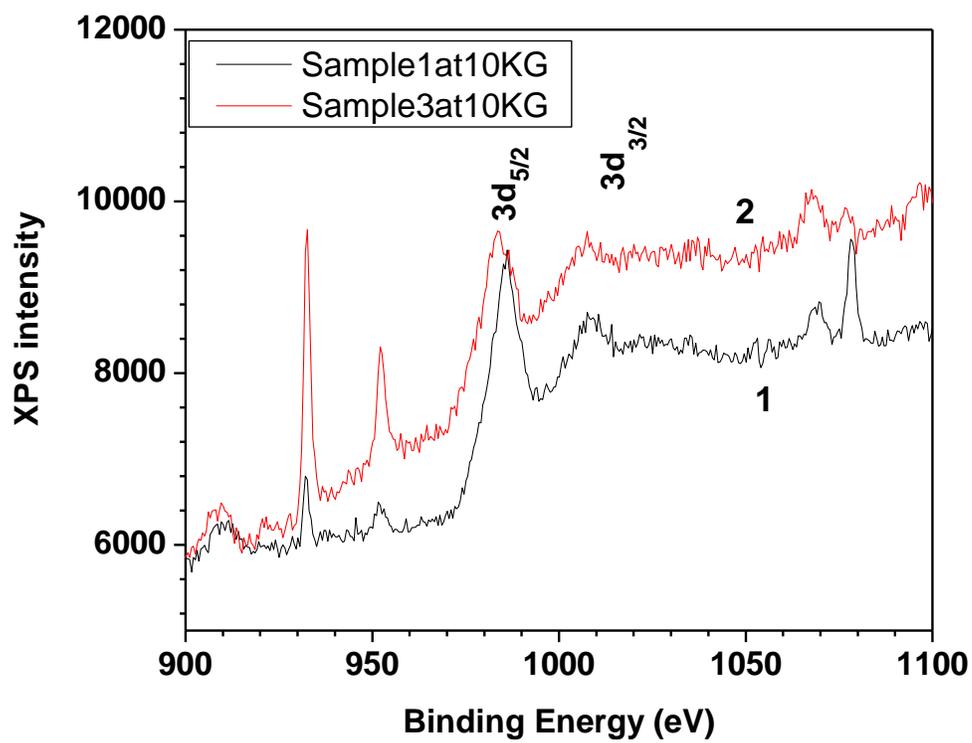

**Fig.- 8**